\documentclass[journal=jctcce,manuscript=article]{achemso}

\SectionNumbersOn

\usepackage[version=3]{mhchem} 

\usepackage{multirow}
\usepackage{listings}
\usepackage{xspace}
\usepackage{float}
\usepackage{booktabs}
\usepackage{braket}
\usepackage{natmove}
\usepackage{todonotes}
\usepackage[flushleft]{threeparttable}

\newcommand{\bigO}[1]{\ensuremath{ \mathcal{O}\left( #1 \right) }}

\newcommand{\rom}[1]{\uppercase\expandafter{\romannumeral #1\relax}}

\newcommand{\acesiii}{ACES \rom{3}}

\newcommand{\etal}{{\em et al.}}

\usepackage[draft,inline,nomargin,index]{fixme}
\fxsetup{theme=color,mode=multiuser}
\FXRegisterAuthor{ed}{efv}{\color{red}EV}
\FXRegisterAuthor{cp}{chp}{\color{blue}CP}

\title{Exploration of Reduced Scaling Formulation of Equation of Motion Coupled-Cluster Singles and Doubles Based on State-Averaged Pair Natural Orbitals}

\author{Chong Peng}

\author{Marjory C. Clement}

\author{Edward F. Valeev}
\affiliation[VT]{Department of Chemistry, Virginia Tech, Blacksburg, VA 24061, USA}
\email{efv@vt.edu}

\begin{document}

\begin{abstract}
	A reduced-complexity variant of equation-of-motion coupled-cluster singles and doubles (EOM-CCSD) method is formulated
	in terms of state-averaged excited state pair natural orbitals (PNO) designed to describe manifolds of excited states.
	State-averaged excited state PNOs for the {\em target} manifold are determined by averaging CIS(D) pair densities over the computational
	manifold. To assess the performance of PNO-EOM-CCSD approach on extended systems
	the new massively parallel canonical EOM-CCSD program has been developed in the Massively Parallel Quantum Chemistry program
	that allows treatment of systems with 50+ atoms using realistic basis sets with 1000+ functions.
	The use of state-averaged PNOs offers several potential advantages relative to the recently proposed state-specific PNOs:
	our approach is robust with respect to root flipping and state degeneracies,
	it is more economical when computing large manifolds of states, and it simplifies evaluation of transition-specific observables such as dipole moments.
	With the PNO truncation threshold of $10^{-7}$, the errors in excitation energies are on average below 0.02 eV for the first six singlet states of 28 organic molecules included in the standard test set of Thiel and co-workers ({\em J. Chem. Phys.} {\bf 2008}, {\em 128}, 134110) with 50-70 state-averaged PNOs per pair.
\end{abstract}

\section{Introduction}
\label{pno_eom_introduction}

Accurate description of electronic spectra of medium ($<100$ atoms) and large ($>100$ atoms) molecular systems has always been a challenge for quantum chemistry. 
The time-dependent density functional theory (TDDFT) is the most popular method for the analysis of excited states due to its computational efficiency, capable of treatment of
systems with hundreds and thousands of atoms. Although TDDFT provides medium accuracy for one-electron excitations,
the accuracy of TDDFT can be limited for certain types of excited states (e.g. Rydberg or charge transfer)\cite{Dreuw2005} and in general its accuracy depends strongly on the density functional\cite{Jacquemin2009}.
In contrast to TDDFT, multiconfiguration/multireference (MR) wave function models, such as MR perturbation theory methods (e.g. complete active space perturbation theory (CASPT2) \cite{Andersson1992} and n-electron valence states perturbation theory (NEVPT2) \cite{Angeli2001}) and MR configuration interaction \cite{Werner1988,Szalay2012}
can recover both static and dynamic electron correlation, can treat multiple electronic states on equal footing,
and attain high accuracy, albeit for rather small systems\cite{Schreiber2008}.
Among the challenges of the MR approaches is the need to select the active space, and the exponential growth of complexity with the size of active space. Although
the latter can be avoided for certain types of systems by numerical approximations such as density matrix renormalization group\cite{Chan2011,Schollwock2005} and other tensor network approaches, the MR methods are generally difficult to use for nonspecialists. Accurate treatment of dynamical electron correlation
in the context of MR methodologies is an ongoing direction of research.

In this work we focus on the treatment of excited states by the coupled-cluster method. The highly robust coupled-cluster hierarchy provides unparalleled accuracy for the ground states by systematically including two-, three- and higher-body correlation effects from a single determinant reference. The CC ansatz can be extended to excited states through the use of the linear-response (LR) theory,\cite{Monkhorst1977} the symmetry-adapted cluster configuration interaction (SAC-CI) method,\cite{Nakatsuji1983,Nakatsuji1979} or
the equation of motion coupled-cluster (EOM-CC) method.\cite{Sekino1984,Stanton1993} However, the high-order scaling of the coupled-cluster methods limits its application to small molecules. Even with truncation to singles and doubles excitations, the excited state CCSD methods still have polynomial scaling with large factor \bigO{N^6} and are constrained to systems containing only 20-30 atoms without access to campus-level or national computing resources.

Recently, the development of {\em reduced scaling} variants of the coupled-cluster methods has been reinvigorated by
Neese's introduction\cite{Neese:2009db} of Pair Natural Orbitals (PNOs) in the context of local correlation formalisms of CC initiated by
Pulay\cite{Pulay1983} and pursued by Werner\cite{Hampel1996,Korona2003} and others.\cite{DanielCrawford2002,Russ2004}
PNOs were originally proposed in the 1960s under the name Pseudo-natural Orbitals. \cite{Edmiston1966,Meyer1971}
Truncation of PNOs significantly reduces the number of unoccupied orbitals while only introducing small errors in correlation energies in post-Hartree-Fock calculations. 
However, the demanding computational cost of the pair-specific integral transformation to the PNO space, which scales $\bigO{N^7}$ if there is no truncation of the PNO space, prevents the development of PNO-based electronic structure theories.
In 2009, Neese  \etal~revived local PNOs (LPNOs) for the CEPA \cite{Neese:2009db} and CCSD \cite{Neese2009a} methods, making use of density fitting approximations to accelerate the integral transformation process. 
It makes large-scale coupled-cluster possible for systems with up to 100 atoms using a small workstation. 
The LPNO approach was improved by imposing block sparsity into cluster operator amplitudes
via domains of projected atomic orbitals (PAOs).\cite{Riplinger2013a,Riplinger2013} The DLPNO-CC method was subsequently
improved via the linear-scaling density fitting\cite{Pinski2015,Riplinger2016} and the introduction of F12 explicit correlation to reduce the basis set error,\cite{Pavosevic2017,Pavosevic2016,Pavosevic2014} culminating in a linear scaling explicitly correlated CCSD(T) method for ground states.
These developments were pursued in parallel by several other groups, with polynomial scaling PNO-CCSD(T) code demonstrated by H\"{a}ttig and co-workers\cite{Schmitz2016}
and a scalable implementation of a linear-scaling PNO-CCSD(T)-F12 demonstrated by Werner and co-workers.\cite{Ma2017}

The key ideas of modern reduced-scaling coupled-cluster methods apply not only to the ground states but also to the excited states.
Two competing visions of how to formulate reduced-scaling excited state methodology have been explored.
H\"attig and Helmich have explored excited-state coupled-cluster methods by introducing the $\bigO{N^4}$ scaling PNO-EOM-CC2 with state-specific PNOs \cite{Helmich2013}, as well as PNO-based CIS(D)\cite{Helmich2011} and ADC(2) \cite{Helmich2014}.  The common ideas to these developments is the use of state-specific PNOs to compress
the cluster operator (computed in the ground state CC equation) and the excited state wave operators for each state, with excited
states computed one at a time. Thus the total number of PNOs grows linearly with the number of excited states.
Recently, Dutta \textit{et al} presented a PNO-based coupled-cluster method for excited states utilizing the similarity-transformed EOM (STEOM) CCSD framework.\cite{Dutta2016,Dutta2018}
In their approach the use of PNOs is limited to the ground state only, with DLPNO-CCSD amplitudes subsequently transformed to the canonical basis and used to
evaluate the bt-PNO-STEOM-CCSD energies of manifolds of states, at a \bigO{N^6} complexity but possible to reach \bigO{N^5} with additional improvements.
However, this approach back-transformed the PNOs to the canonical space in the equation of motion CCSD calculations, thus limiting the size of the system can be considered.
We should also note that the use of local correlation ideas (namely, PAO domains) without the PNO-style compression,
have been explored in the context of coupled-cluster methods for excitation energies,
such as local EOM-CC2\cite{Kats2006} and local EOM-CCSD \cite{DanielCrawford2002, Korona2003}.

In this work, we present a PNO-based approach suitable for robust treatment of manifolds of excited states with the
EOM-CCSD methods.
The key idea is to use state-averaged PNOs similar to those used in the ground-state PNO coupled-cluster methods through state-averaged guess pair densities averaged over the target excited state manifold. 
To quickly explore the performance of our approach we simulated it using a massively parallel EOM-CCSD implementation.
The new massively parallel EOM-CCSD was implemented in the Massively Parallel Quantum Chemistry (MPQC) package \cite{mpqc4} using the TiledArray\cite{tiledarray} framework, based on the ground-state CCSD implementation described previously.\cite{Peng2016}
The new implementation exhibits good strong-scaling parallel performance and allows the calculation of excitation energy for systems with more than 50 atoms and more than 1000 basis functions; this is crucial to the exploration of the state-averaged PNO ansatz for systems of realistic size.
In Section \ref{pno_eom_methods}, the theory and implementation of state-averaged PNOs are discussed. 
Section \ref{pno_eom_computational_detail} describes the computational details as well as the computing resources used.
Section \ref{pno_eom_result} demonstrates the performance of the parallel EOM-CCSD code and the accuracy of state-averaged PNOs.

\section{Methods}
\label{pno_eom_methods}
The coupled-cluster ground-state wave function,
\begin{equation}
\Psi_{(0)} \equiv  e^{\hat{T}} \ket{0},
\end{equation}
where the $\ket{0}$ stands for the zeroth order reference wave function (usually a Hartree-Fock determinant),
is determined by projection of the Schr\"odinger equation against excited determinants
\begin{align}
0 & = \bra{\overline{^a_i}} \bar{H} \ket{0},  \\
0 & = \bra{\overline{^{ab}_{ij}}} \bar{H} \ket{0}. 
\end{align}
with $\bar{H} \equiv e^{-\hat{T}} H e^{\hat{T}}$ the usual similarity-transformed Hamiltonian.
Within the equation of motion coupled-cluster method\cite{Sekino1984,Stanton1993} $k$th excited-state wave function
is obtained in a CI fashion, by the action of a linear excitation operator acting on the ground-state CC wave function:
\begin{equation}
\Psi_{(k)} \equiv  \hat{R}_{(k)} e^{\hat{T}} \ket{0},
\end{equation}
$\hat{R}_{(k)}$ and the corresponding energies $E_{(k)}$ are obtained by diagonalizing
the similarity-transformed Hamiltonian:
\begin{equation}
\bar{H} \hat{R}_{(k)} \ket{0} = E_{(k)} \hat{R}_{(k)} \ket{0}.
\end{equation} 
In practice the ground and excited states are represented in terms of single and double excitations only:
\begin{align}
\hat{T} & \overset{\text{CCSD}}{\equiv}  \hat{T}_1 + \hat{T}_2, \\
\hat{T}_{1} & \equiv T^{i}_{a} E_{i}^{a} , \\
\hat{T}_{2} & \equiv \frac{1}{2} T^{ij}_{ab} E_{ij}^{ab}. \\
\hat{R}_{(k)} & \overset{\text{CCSD}}{\equiv} \delta_{k0} + \hat{R}_{1(k)} + \hat{R}_{2(k)}, \\
\hat{R}_{1(k)} & \equiv B^{i}_{a(k)} E_{i}^{a} , \\
\hat{R}_{2(k)} & \equiv \frac{1}{2} B^{ij}_{ab(k)} E_{ij}^{ab}.
\end{align}
The storage and operation costs of the CCSD and EOM-CCSD methods are $\bigO{N^4}$ and $\bigO{N^6}$, respectively.

Two-body amplitude tensors, $T^{ij}_{ab}$ and $B^{ij}_{ab}$, are efficiently rank-compressed by transforming each $ij$ block into the $ij$-specific subspace.
For the ground-state amplitudes the optimal pair-specific subspaces are robustly approximated by the truncated singular subspace of the corresponding ground-state pair densities, $\textbf{D}_{(0)}^{ij}$, computed from guess amplitudes:
\begin{align}
\textbf{D}_{(0)}^{ij} = \frac{2}{1 + \delta_{ij}} (\textbf{T}^{ij} \tilde{\textbf{T}}^{ij \dagger} + \textbf{T}^{ij \dagger} \tilde{\textbf{T}}^{ij} ),
\end{align}
where $\left( \textbf{T}^{ij} \right)_{ab} \equiv T^{ij}_{ab}$ and $\tilde{\textbf{T}}^{ij} \equiv 2 \textbf{T}^{ij} - \textbf{T}^{ij \dagger}$.
(Semi)canonical MP1 amplitudes,
\begin{equation}
(T^{(1)})^{ij}_{ab} \equiv \frac{G^{ij}_{ab}}{f_{i}^{i} + f_{j}^{j} - f_{a}^{a} - f_{b}^{b}},
\label{eq:T1}
\end{equation}
are typically used as the guess \cite{Neese2009a} (In Eq. \eqref{eq:T1} $f$ are the matrix elements of the Fock operator, and $G$ stands for the two-electron integral).
PNOs are the basis for the singular subspace of $\textbf{D}_{(0)}^{ij}$ ; they are obtained by solving the eigensystem:
\begin{equation}
\textbf{D}_{(0)}^{ij} \textbf{U}_{(0)}^{ij} = \textbf{U}_{(0)}^{ij} \textbf{n}_{(0)}^{ij}.
\end{equation}
where $\textbf{n}_{(0)}^{ij}$ are the PNO occupation numbers.
PNOs with occupation numbers less than user-provided threshold $T_\text{CutPNO}$ are omitted, hence the number of PNOs per pair is independent of the system size (i.e. $\bigO{1}$).
One-body amplitudes, $T^{i}_{a}$ and $B^{i}_{a}$, are compressed similarly to the two-body counterparts by transforming into the basis of orbital-specific virtuals (OSVs)\cite{Yang2011}. OSVs are traditionally defined to be identical to the PNOs of diagonal pairs but truncated according to a different threshold, $T_\text{CutOSV}$.

As pointed out by H\"{a}ttig and Helmich, \cite{Helmich2011} the optimal singular subspaces for the ground and excited state amplitudes differ; as a result, the PNOs and OSVs must be constructed separately for the ground and excited states.
H\"attig and Helmich proposed the use of state-specific PNOs, where the PNOs for each state are constructed using CIS(D) doubles amplitudes with respect to that state:
\cite{Helmich2013}
\begin{equation}
B^{ij}_{ab (k)} = \frac{K^{ij}_{ab (k)}}{\omega_{(k)} + f_{i}^{i} + f_{j}^{j} - f_{a}^{a} - f_{b}^{b}},
\label{eq:cis_d}
\end{equation}
\begin{equation}
K^{ij}_{ab (k)} = B_{c (k)}^{i} G_{ab}^{cj} + B_{c (k)}^{j} G_{ab}^{ic} - B_{a (k)}^{l} G_{lb}^{ij} - B_{b (k)}^{l} G^{ij}_{al}  ,
\end{equation}
where $\textbf{B}^{i}_{a (k)}$ and $\textbf{B}^{ij}_{ab (k)}$ are the CIS singles amplitudes and CIS(D) doubles amplitudes for excited state $k$, and $\omega_{(k)}$ is the CIS excitation energy. 
The state-specific PNOs for excited states can be obtained from the state-specific pair density using the CIS(D) doubles amplitudes similar to the approach used in ground state:
\begin{equation}
\textbf{D}^{ij}_{(k)} = \frac{2}{1 + \delta_{ij}} (\textbf{B}^{ij}_{(k)} \tilde{\textbf{B}}^{ij \dagger}_{(k)} + \textbf{B}^{ij \dagger}_{(k)} \tilde{\textbf{B}}^{ij}_{(k)} ),
\end{equation}
Such definition of excited state PNOs yields good accuracy in the context of PNO-EOM-CC2 method.\cite{Helmich2013}
However, there are several factors that prompted us to look beyond the state-specific PNOs. First, and foremost, the cost of PNO construction and integral transformation grow linearly with the number of states. This is particularly notable since the cost of PNO-based methods is often dominated by the cost of the integral transformation, even when domain approximations are employed.\cite{Pinski2015,Riplinger2016} Second, state-specific PNOs make it difficult to deal with degenerate state manifolds (which ideally need to be expressed in the same basis).
Lastly, the use of state-specific PNOs increases the complexity of formalism and implementation.

Thus we decided to investigate PNO-EOM-CCSD that uses one set of PNOs for all excited states, in particular, we propose to use the {\em state-averaged} PNOs.
The state-averaged PNOs are defined as the eigenvectors of averaged pair densities over an $N$-state manifold:
\begin{equation}
\label{eq:eom:sa_density}
\textbf{D}^{ij} = \frac{1}{N} \sum_{k}^{N} \textbf{D}^{ij}_{(k)}.
\end{equation}
(State-averaged OSVs will be defined in this work the PNOs of the diagonal pairs, in complete analogy with the construction of the ground-state OSVs). 

Although the work is underway in our group to develop a production implementation of reduced-scaling CC, here our goal is more modest: we aim to evaluate
the proposed state-averaged PNO formulation in the context of EOM-CCSD. Hence we initially implemented
a simulation for PNO-EOM-CCSD based on a newly-developed massively parallel canonical (i.e., \bigO{N^6}) EOM-CCSD program in the MPQC code.
Note that simulation has been used previously for initial evaluation of locally-correlated PAO-based EOM-CCSD by Russ and Crawford\cite{DanielCrawford2002,Russ2004} and by Korona and Werner \cite{Korona2003}. Werner \etal~and Crawford \etal~also used simulation to compare PAO-, OSV-, and PNO-based formulations of CCSD.\cite{Krause2012,McAlexander2016}
It should also be noted that H\"attig and Helmich have demonstrated {\em production}-level \bigO{N^4} PNO-EOM-CC2 methods,\cite{Helmich2013} but PNO-EOM-CCSD has not yet been reported at the time of writing this manuscript.

The canonical closed-shell EOM-CCSD program in MPQC was implemented on top of the TiledArray tensor framework following the formalism of Bartlett and Stanton \cite{Stanton1993}. The implementation
details generally follow the ground-state explicitly correlated CCSD implementation reported previously.\cite{Peng2016} 
All amplitudes and intermediates are distributed in memory, and contractions are evaluated using the communication-optimal implementation of the distributed-memory scalable universal matrix multiplication algorithm (SUMMA) implemented in TiledArray.\cite{Calvin2015a,Calvin2015b} 
Similarly to the ground-state CCSD, the largest intermediate needed to compute in the EOM-CCSD is the $W_{ab}^{cd}$ term with four virtual indices:
\begin{equation}
W_{ab}^{cd} \equiv G_{ab}^{cd} - T^{i}_{b} G_{ai}^{cd} - T^{i}_{a}G_{ib}^{cd} + (T_{ab}^{ij} + T_{a}^{i}T_{b}^{j})G_{ij}^{cd}.
\end{equation}
When contracting with $B_{ab}^{ij}$, this intermediate can be avoided through a back-transformed intermediate:
\begin{align}
W_{ab}^{cd}B_{cd}^{ij} &= X^{ij}_{\rho \sigma} C^{\rho}_{a} C^{\sigma}_{b} - X^{ij}_{\sigma \rho} C^{\rho}_{k} C^{\sigma}_{a} T_{b}^{k} - X^{ij}_{\rho \sigma} C^{\rho}_{k} C^{\sigma}_{b} T_{a}^{k}
+ X^{ij}_{\rho \sigma} C^{\rho}_{k} C^{\sigma}_{l} (T_{ab}^{kl} + T_{a}^{k}T_{b}^{l}) ,\\    
X^{ij}_{\rho \sigma} &= (B_{cd}^{ij} C_{\mu}^{c} C_{\nu}^{d}) G_{\rho  \sigma}^{\mu \nu}.
\end{align}
Computing intermediate $X$ requires evaluating atomic two-electron integral on the fly.
In this way, the storage requirements of the EOM-CCSD program have been reduced, allowing us to carry out calculations on systems with over 1000 basis functions.
The same technique has been used by Ku{\'{s}} \etal{} in \acesiii. \cite{Kus2009} 

The ground-state PNO-CCSD simulation was implemented with modification to the Jacobi update in the following manner:
\begin{enumerate}
	\item After the CCSD amplitude residuals $\textbf{R}_{1}$ and $\textbf{R}_{2}$ are computed, $\textbf{R}_{1}$ is transformed into a semi-canonical OSV basis and $\textbf{R}_{2}$ is transformed into a semi-canonical PNO basis:
	\begin{equation}
	\bar{\textbf{R}}^{i} = \textbf{U}^{i \dagger} \textbf{R}^{i},
	\end{equation}
	\begin{equation}
	\bar{\textbf{R}}^{ij} = \textbf{U}^{ij \dagger} \textbf{R}^{ij} \textbf{U}^{ij},
	\end{equation}
	where $\textbf{R}^{i}$/$\textbf{R}^{ij}$ are the corresponding orbital/pair blocks of $\textbf{R}_{1}$/$\textbf{R}_{2}$ residuals, and $\textbf{U}^{i}$/$\textbf{U}^{ij}$ are the ground-state OSV/PNO bases.
	\item The residuals are updated through a Jacobi update in the OSV and PNO space:
	\begin{equation}
	\bar{\Delta}^{i}_{a_{i}} = \frac{\bar{R}^{i}_{a_{i}}} {f_{i}^{i}  - \bar{f}_{a_{i}}^{a_{i}} },
	\end{equation}
	\begin{equation}
	\bar{\Delta}^{ij}_{a_{ij}b_{ij}} = \frac{\bar{R}^{ij}_{a_{ij}b_{ij}}} {f_{i}^{i} + f_{j}^{j} - \bar{f}_{a_{ij}}^{a_{ij}} - \bar{f}_{b_{ij}}^{b_{ij}} },
	\end{equation}
	where $a_{i}$ and $a_{ij}$ are unoccupied orbitals in the truncated OSV and PNO basis, respectively.
	\item The updated residuals are extrapolated with DIIS and back-transformed into the canonical basis:
	\begin{equation}
	\mathbf{\Delta}^{i} = \textbf{U}^{i} \bar{\mathbf{\Delta}}^{i},
	\end{equation}
	\begin{equation}
	\mathbf{\Delta}^{ij} = \textbf{U}^{ij} \bar{\mathbf{\Delta}}^{ij} \textbf{U}^{ij \dagger}.
	\end{equation}
	\item The new CCSD amplitudes are formed as an update to the current amplitudes:
	\begin{equation}
	\textbf{T}^{i}_{n+1} = \textbf{T}^{i}_{n} + \mathbf{\Delta}^{i} ,
	\end{equation}
	\begin{equation}
	\textbf{T}^{ij}_{n+1} = \textbf{T}^{ij}_{n} + \mathbf{\Delta}^{ij},
	\end{equation}
	where $n$ stands for the number of current iteration.
	\item The CCSD residuals are recomputed using the new amplitudes, and this process is repeated from step 1 until convergences is reached.
\end{enumerate}

Similarly, the state-averaged PNO simulation in EOM-CCSD can be done with modification in the Davidson solver:
\begin{enumerate}
	\item The residuals produced by the Davidson algorithm are transformed into the OSV and PNO bases:
	\begin{equation}
	\bar{\textbf{R}}^{i}_{(k)} = \textbf{U}^{i \dagger} \textbf{R}^{i}_{(k)},
	\end{equation}
	\begin{equation}
	\bar{\textbf{R}}^{ij}_{(k)} = \textbf{U}^{ij \dagger} \textbf{R}^{ij}_{(k)} \textbf{U}^{ij}.
	\end{equation}
	
	\item A preconditioner is applied to the residuals in the OSV and PNO spaces:
	\begin{equation}
	\bar{B}^{i}_{a_{i}(k)} = \frac{\bar{R}^{i}_{a_{i}(k)}} { \omega_{(k)} + f_{i}^{i}  - \bar{f}_{a_{i}}^{a_{i}} },
	\end{equation}
	\begin{equation}
	\bar{B}^{ij}_{a_{ij}b_{ij}(k)} = \frac{\bar{R}^{ij}_{a_{ij}b_{ij}(k)}} { \omega_{(k)} + f_{i}^{i} + f_{j}^{j} - \bar{f}_{a_{ij}}^{a_{ij}} -  \bar{f}_{b_{ij}}^{b_{ij}} } ,
	\end{equation}
	where $\omega_{(k)}$ is the eigenvalue of state $k$.
	
	\item The updated trial vectors are projected back into the canonical space:
	\begin{equation}
	\textbf{B}^{i}_{(k)} = \textbf{U}^{i} \bar{\textbf{B}}^{i}_{(k)} ,
	\end{equation}
	\begin{equation}
	\textbf{B}^{ij}_{(k)} = \textbf{U}^{ij} \bar{\textbf{B}}^{ij}_{(k)} \textbf{U}^{ij \dagger} .
	\end{equation}
	
	\item The new trial vectors are added to the next iteration of the Davidson algorithm to update the subspace, and the process is continued from step 1 until it reached convergence. 
	
\end{enumerate}

\section{Computational Details}
\label{pno_eom_computational_detail}
The canonical EOM-CCSD code was implemented and tested in the developmental version of the MPQC program.\cite{mpqc4}
All computations were performed on a commodity cluster at Virginia Tech, each node of which has two
Intel Xeon E5-2670 CPUs (332 GFLOPS) and 64 GB of RAM.
MPQC was compiled using GCC 5.3.0 with Intel MPI 5.0 and the serial Intel MKL version 11.2.3. 
All computations launched 1 MPI process per node with 16 threads per MPI process, with the orbital block size set to 20.

The state-averaged PNO-EOM-CCSD simulation code was also implemented in MPQC. 
To simplify the definition of PNOs and OSVs truncation, we set $T_{\textmd{CutOSV}}$=$T_{\textmd{CutPNO}}/10$ for both ground and excited states.
Neither domain nor weak pair approximations were utilized.
The occupied MOs were localized in all calculations via the Foster-Boys algorithm \cite{Foster1960, Boys1960}.
The density-fitting (resolution-of-identify) approximation \cite{Feyereisen1993, Vahtras1993} and frozen core approximation were used for all the calculations performed in this work.
We have used the cc-pVTZ \cite{Dunning1989} and aug-cc-pVD/TZ \cite{Kendall1992} atomic orbital basis sets in our calculations, with the corresponding auxiliary basis sets cc-pVTZ-RI and aug-cc-pVD/ TZ-RI \cite{Weigend2002} for density-fitting. 
In Section \ref{pno_eom_result}, the geometries of the methylated uracil dimer with water and the phenolate form of the anionic chromophore of the photoactive yellow protein were obtained from Ref. \citenum{Epifanovsky2013}.
The structure of 11-cis-retinal protonated Schiff base was obtained from Ref. \citenum{Dutta2016}.
The structures of benzonitrile and acetamide were obtained from Ref. \citenum{Schreiber2008}.
In Section \ref{section:error_analysis}, a total number of 10 excited-states were computed for  the benchmark dataset of 28 organic molecules by Thiel \cite{Schreiber2008}.

\section{Results \& Discussion}
\label{pno_eom_result}

\subsection{Parallel Performance of EOM-CCSD}
The new canonical EOM-CCSD code can attain high efficiency and good parallel scalability as illustrated in Fig. 1 and Fig. 2 for realistic computations (with aug-cc-pVDZ and cc-pVTZ basis sets) on excited-states of the methylated uracil dimer with water and 11-cis-retinal protonated Schiff base, respectively. 
\begin{figure}
	\centering
	\includegraphics[width=0.7\linewidth]{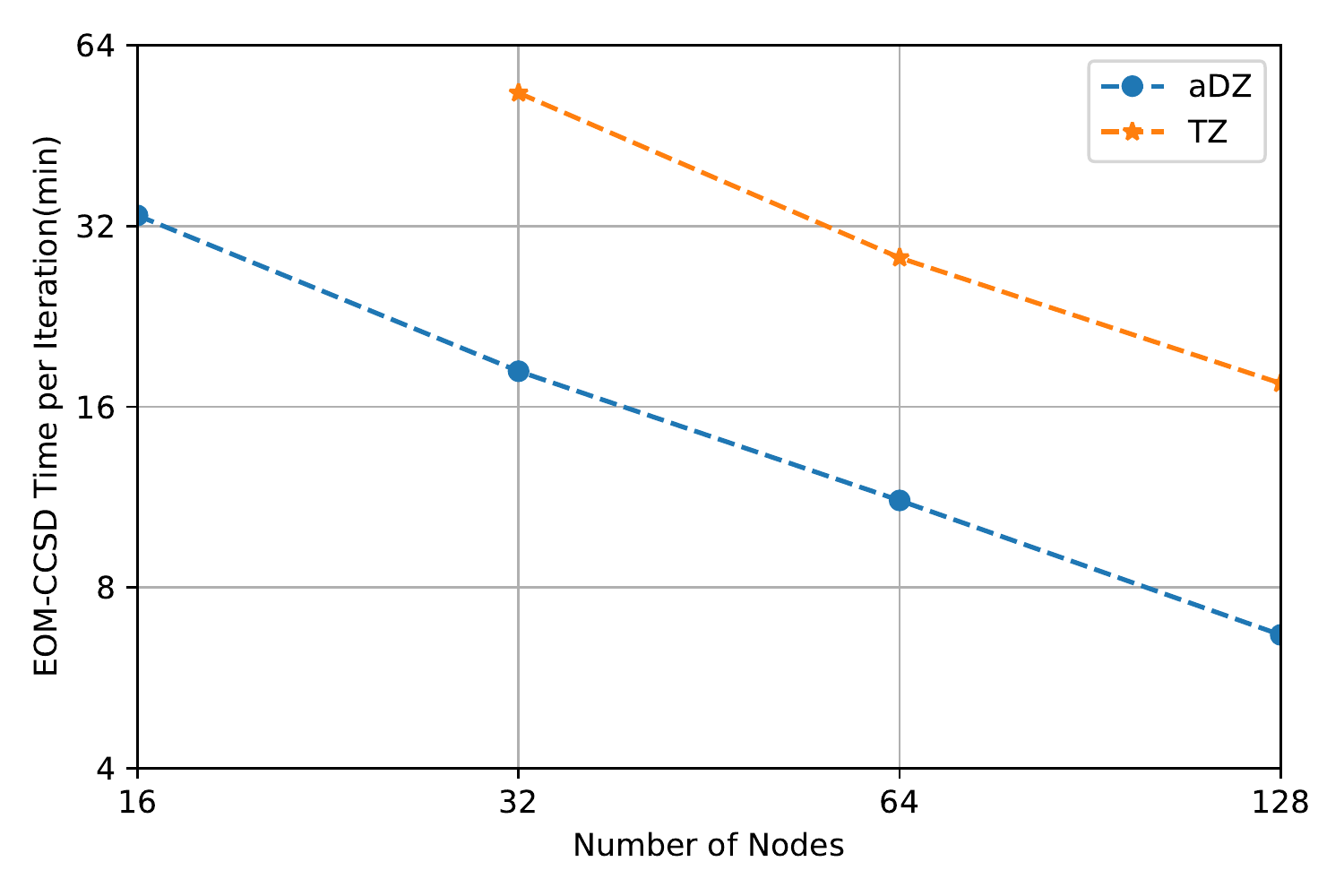}
	\caption{Parallel performance of EOM-CCSD on 4 states of the methylated uracil dimer with water (39 atoms) with the aug-cc-pVDZ (645) and cc-pVTZ (882) basis sets}
	\label{fig:eom:mU_h2o}
\end{figure}
\begin{figure}
	\centering
	\includegraphics[width=0.7\linewidth]{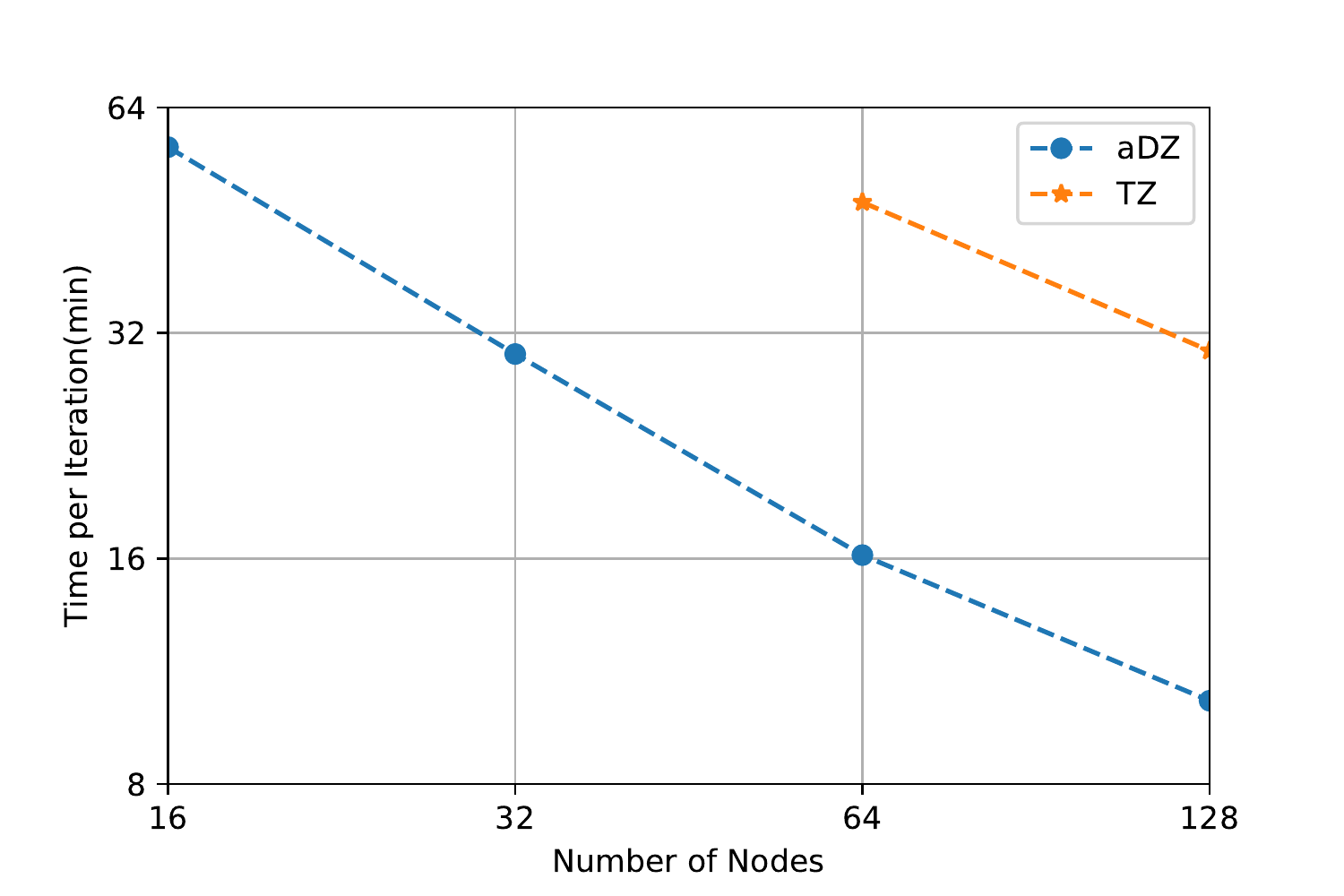}
	\caption{Parallel performance of EOM-CCSD on 4 states of the 11-cis-retinal protonated Schiff base (51 atoms) with the aug-cc-pVDZ (753) and cc-pVTZ (1050) basis sets}
	\label{fig:eom:cis_retinal}
\end{figure}
The data in Fig. \ref{fig:eom:mU_h2o} corresponds to a $\sim5$ speedup from 16 to 128 nodes
with the aug-cc-pVDZ basis and a $\sim3$ speedup from 32 to 128 nodes with the cc-pVTZ basis.
The data in Fig. \ref{fig:eom:cis_retinal} corresponds to a $\sim5.5$ speedup from 16 to 128 nodes with the aug-cc-pVDZ basis and a $\sim1.6$ speedup from 64 to 128 nodes with the cc-pVTZ basis.
The demonstrated strong scaling is not as impressive as that of the ground-state CCSD program,\cite{Peng2016} but additional improvements are planned.
The performance of our code is already sufficient to be able to treat multiple states of a system with 50-100 atoms and 1000-1500 basis functions.

\subsection{Accuracy of State-Averaged PNOs}

To quantify the performance of state-averaged PNOs we computed errors in excitation energies relative to the canonical EOM-CCSD values
introduced by the truncation of PNOs (and the corresponding truncation of OSVs).
Table \ref{table:eom:sa_pno} lists the PNO truncation errors for benzonitrile in the cc-pVTZ basis for a fixed value of the $T_{\textmd{CutPNO}}$ parameter, as a function of
the number of computed states.

\begin{table}[!h]
	\caption{Truncation errors in excitation energy (eV) of benzonitrile (cc-pVTZ,V=283 \textsuperscript{\emph{a}}) with respect to total number of states at $T_{\textmd{CutPNO}}$=$10^{-8}$}
	\label{table:eom:sa_pno}
	\begin{center}
		\begin{tabular}{lrrrrrrr}
			\toprule
			nStates &      1  &      2  &      4  &      6  &      8  &      10 &      20 \\
			ESnPNO   &      63 &     79  &     91  &     95  &    101  &     104 &     133 \\
			\midrule
			$S_1$     &  0.0072 &  0.0016 &  0.0014 &  0.0003 & -0.0001 & -0.0001 & -0.0011 \\
			$S_2$     &         &  0.0043 &  0.0023 &  0.0008 &  0.0005 &  0.0004 & -0.0005 \\
			$S_3$     &         &         &  0.0235 &  0.0039 &  0.0019 &  0.0019 &  0.0005 \\
			$S_4$     &         &         &  0.0263 &  0.0022 &  0.0022 &  0.0022 &  0.0009 \\
			$S_5$     &         &         &         &  0.0161 &  0.0173 &  0.0171 &  0.0040 \\
			$S_6$     &         &         &         &  0.0069 &  0.0049 &  0.0042 &  0.0004 \\
			$S_7$     &         &         &         &         &  0.0876 &  0.0235 &  0.0050 \\
			$S_8$     &         &         &         &         &  0.1834 &  0.0557 &  0.0012 \\
			$S_9$     &         &         &         &         &         &  0.0201 & -0.0001 \\
			$S_{10}$  &         &         &         &         &         &  0.0058 &  0.0044 \\
			MAE \textsuperscript{\emph{b}}   &  0.0072 &  0.0030 &  0.0134 &  0.0050 &  0.0372 &  0.0131 &  0.0018 \\
			MAX \textsuperscript{\emph{c}}  &  0.0072 &  0.0043 &  0.0263 &  0.0161 &  0.1834 &  0.0557 &  0.0050 \\
			\bottomrule
		\end{tabular}
	\end{center}
	\begin{tablenotes}
		\item[1] \textsuperscript{\emph{a}} Total number of unoccupied orbitals 
		\item[2] \textsuperscript{\emph{b}} Mean absolute error
		\item[3] \textsuperscript{\emph{c}} Max absolute error 
	\end{tablenotes}

\end{table}

As expected, the average number of excited-state PNOs (ESnPNO) increases with the total number of states. However, the rate of increase is rather modest: raising the number of states from 1 to 20 increases the number of PNOs only by a factor of $\sim2$. Clearly, the total number of state-averaged PNOs grows with the number of states far slower than the linear growth of the total number of state-specific PNOs used by H\"attig and co-workers. This is not entirely surprising; since the low-energy states in molecules to zeroth order have many occupied orbitals in common; correlation effects will be largely similar among the states. Clearly, the use of state-averaged PNOs should offer substantial savings in the costs of the integral transformation.
The errors in excitation energies also decrease as the total number of states increases because of the concomitant increase in the number of PNOs.
On average our approach to PNO construction is rather accurate: the mean absolute errors are below 0.02 eV for all cases except nStates=8, which has a mean absolute error of 0.037 eV. These errors are small relative to the average accuracy of the EOM-CCSD model even for states with single excitation character.

Note that the mean absolute errors do not smoothly decrease as nStates increases. This is correlated with sporadic increases in the maximum absolute errors as the number of states is increased, such as in the case of states 3 and 4 when nStates is at 4 and states 7 and 8 when nStates is at 8.
However, these errors are significantly reduced when nStates is increased to 6 and 10, respectively.
This indicates that the highest excited states in the computed manifold sometimes have more significant errors with state-averaged PNOs, which can be observed from the nStates = 4,8 data. 
The reason for this behavior is that the composition of $N$ lowest EOM-CCSD states may not be similar to that of CIS, either due to pure root flipping or, more generally, due to nonperturbative effects of dynamical correlation on the excited state character and ordering.
Analysis of the excited states in this example suggests that CIS states 3, 4, 5, and 6 become EOM-CCSD states 5, 6, 3, and 4. Therefore accurate description
of EOM-CCSD states 3 and 4 will require including pair densities from CIS(D) states 5 and 6. This is not a serious issue since in excited state computations
to increase the probability that $N$ lowest-energy target states have been reproduced is to compute $M > N$ states.
Hence, a slightly larger error in a few of the highest excited states would not be an issue since typically the number of {\em computed} states is always
greater than the number of {\em target} states.

Lastly, note that the $T_{\textmd{CutPNO}}$ threshold is kept constant in Table 1. Therefore the averaged errors decrease as the number of states increases, at the cost of increasing the average number of state-averaged PNOs per pair. Clearly, if we wanted to keep the average error per state constant we could loosen the $T_{\textmd{CutPNO}}$ threshold as the number of states is increased. This would further alleviate the modest increase of the total number of PNOs with the number of states. Dependence of the error
on $T_{\textmd{CutPNO}}$ will be examined next.

Tables \ref{table:eom:pypb_gs}, \ref{table:eom:pypb_es} and \ref{table:eom:pypb_gses} illustrate the correlation between  the $T_{\textmd{CutPNO}}$ parameter and the errors in the excitation energies of the 4 lowest singlet excited states of the phenolate form of the anionic chromophore of the photoactive yellow protein (PYPb). Since $T_{\textmd{CutPNO}}$
affects the EOM-CCSD excitation energies through both ground-state ($\hat{T}$) and excited-state ($\hat{R}$) operators, we examined its effects separately on the ground-state cluster operators only (Table \ref{table:eom:pypb_gs}), excited-state operators (Table \ref{table:eom:pypb_es}) and both (Table \ref{table:eom:pypb_gses}).
\begin{table}[!h]
	\caption{Truncation error in excitation energy (eV) of PYPb (aug-cc-pVDZ, V=296 \textsuperscript{\emph{a}}) by only truncating the ground-state PNOs}
	\label{table:eom:pypb_gs}
	\begin{center}
		\begin{tabular}{lrrrrr}
			\toprule
			$T_{\textmd{CutPNO}}$ &    $10^{-6}$&     $10^{-7}$ &     $10^{-8}$ &     $10^{-9}$ &    $10^{-10}$ \\
			GSnPNO &          13  &         25    &        45     &       74      &      110      \\
			\midrule
			$S_1$     &       -0.0872 &       -0.0289 &       -0.0092 &       -0.0030 &       -0.0011 \\
			$S_2$     &       -0.0892 &       -0.0301 &       -0.0098 &       -0.0033 &       -0.0012 \\
			$S_3$     &       -0.0855 &       -0.0287 &       -0.0093 &       -0.0031 &       -0.0012 \\
			$S_4$     &       -0.0875 &       -0.0296 &       -0.0098 &       -0.0033 &       -0.0013 \\
			\bottomrule
		\end{tabular}
	\end{center}    
	\begin{tablenotes}
		\item[1] \textsuperscript{\emph{a}} Total number of unoccupied orbitals 
	\end{tablenotes}
\end{table}
\begin{table}[!h]
	\caption{Truncation error in excitation energy (eV) of PYPb (aug-cc-pVDZ, V=296 \textsuperscript{\emph{a}}) by only truncating the excited-state PNOs}
	\label{table:eom:pypb_es}
	\begin{center}
		\begin{tabular}{lrrrrr}
			\toprule
			$T_{\textmd{CutPNO}}$ &    $10^{-6}$&     $10^{-7}$ &     $10^{-8}$ &     $10^{-9}$ &    $10^{-10}$ \\
			ESnPNO &          6   &         15    &        32     &       58      &       94      \\
			\midrule
			$S_1$     &        0.1323 &        0.0564 &        0.0211 &        0.0053 &        0.0015 \\
			$S_2$     &        0.1215 &        0.0525 &        0.0197 &        0.0047 &        0.0011 \\
			$S_3$     &        0.1354 &        0.0635 &        0.0256 &        0.0078 &        0.0024 \\
			$S_4$     &        0.1296 &        0.0611 &        0.0261 &        0.0092 &        0.0035 \\
			\bottomrule
		\end{tabular}
	\end{center}
	\begin{tablenotes}
		\item[1] \textsuperscript{\emph{a}} Total number of unoccupied orbitals 
	\end{tablenotes}   
\end{table}
\begin{table}[!h]
	\caption{Truncation error in excitation energy (eV) of PYPb (aug-cc-pVDZ, V=296 \textsuperscript{\emph{a}}) by truncating both the ground and excited-state PNOs}
	\label{table:eom:pypb_gses}
	\begin{center}
		\begin{tabular}{lrrrrr}
			\toprule
			$T_{\textmd{CutPNO}}$ &    $10^{-6}$&     $10^{-7}$ &     $10^{-8}$ &     $10^{-9}$ &    $10^{-10}$ \\
			GSnPNO &          13  &         25    &        45     &       74      &      110      \\
			ESnPNO &          6   &         15    &        32     &       58      &       94      \\
			\midrule
			$S_1$     &        0.0457 &        0.0276 &        0.0119 &        0.0022 &        0.0003 \\
			$S_2$     &        0.0332 &        0.0226 &        0.0098 &        0.0013 &       -0.0002 \\
			$S_3$     &        0.0500 &        0.0348 &        0.0163 &        0.0046 &        0.0012 \\
			$S_4$     &        0.0422 &        0.0315 &        0.0163 &        0.0058 &        0.0022 \\
			\bottomrule
		\end{tabular}
	\end{center}    
	\begin{tablenotes}
		\item[1] \textsuperscript{\emph{a}} Total number of unoccupied orbitals 
	\end{tablenotes}
\end{table}
As expected (see Table \ref{table:eom:pypb_gs}) truncating the ground-state PNOs only lowers the excitation energies (the errors in excited states are all negative) since the energy of the ground state becomes higher due to a decrease in the amount of correlation energy that is recovered. 
Similarly, truncating the excited-state PNOs raises the excitation energy since the calculated energy of the excited states is now higher as a result of recovering less of the correlation energy (Table \ref{table:eom:pypb_es}).
When both the ground and excited state PNOs are truncated, the opposite signs of the two sources of error partially cancel each other out, leading to smaller errors in the excitation energy, as can be seen in Table \ref{table:eom:pypb_gses}. However, this error cancellation may lead to occasional non-monotonic convergence.

\subsection{Error Analysis}
\label{section:error_analysis}
\begin{figure}
	\centering
	\includegraphics[width=0.7\linewidth]{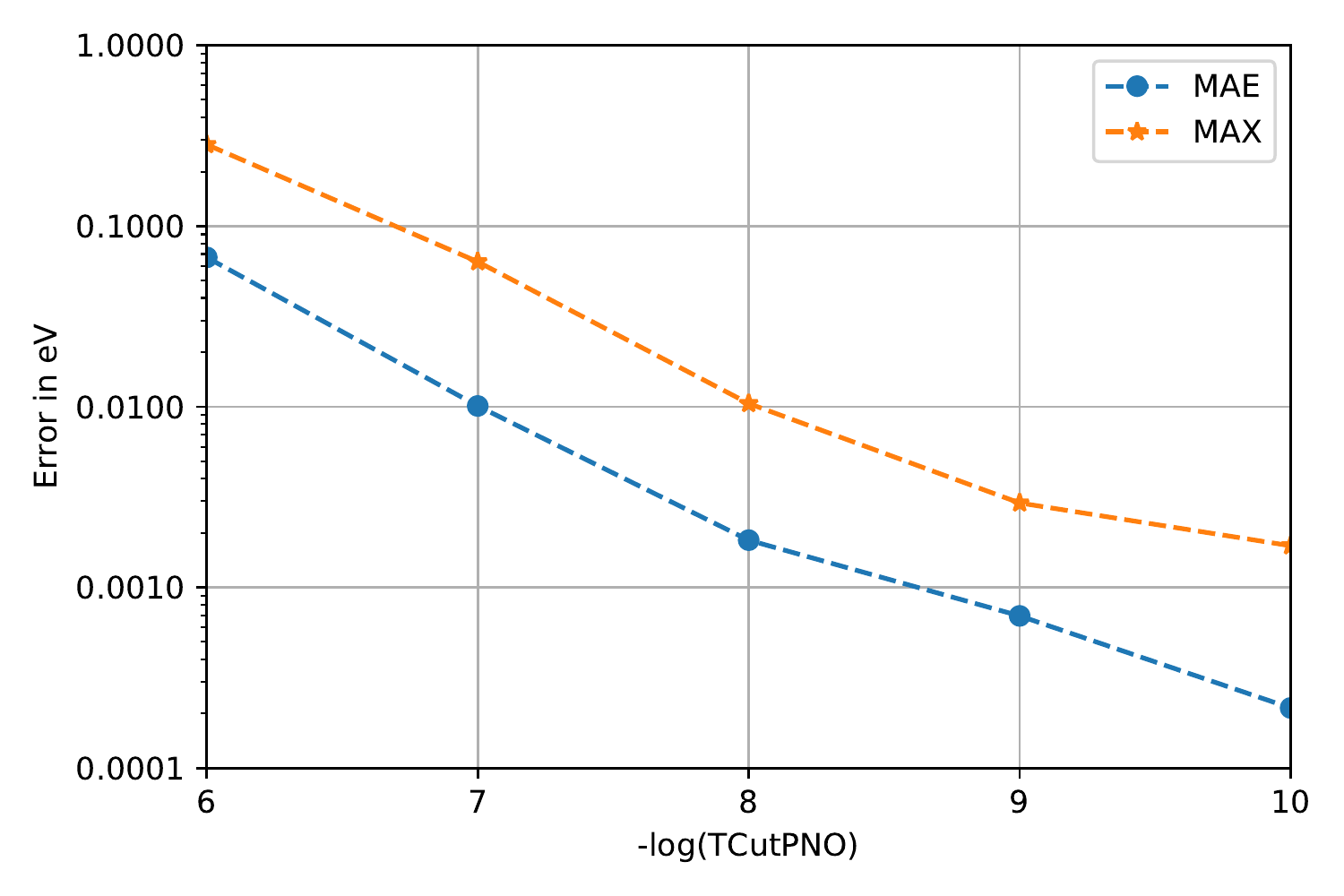}
	\caption{Mean absolute (MAE) and maximum (MAX) PNO truncation errors (in eV) of PNO-EOM-CCSD/cc-pVTZ excitation energies for the 28-molecule benchmark set.}
	\label{fig:eom:benchmark_tz_ee}
\end{figure}

\begin{figure}
	\centering
	\includegraphics[width=0.7\linewidth]{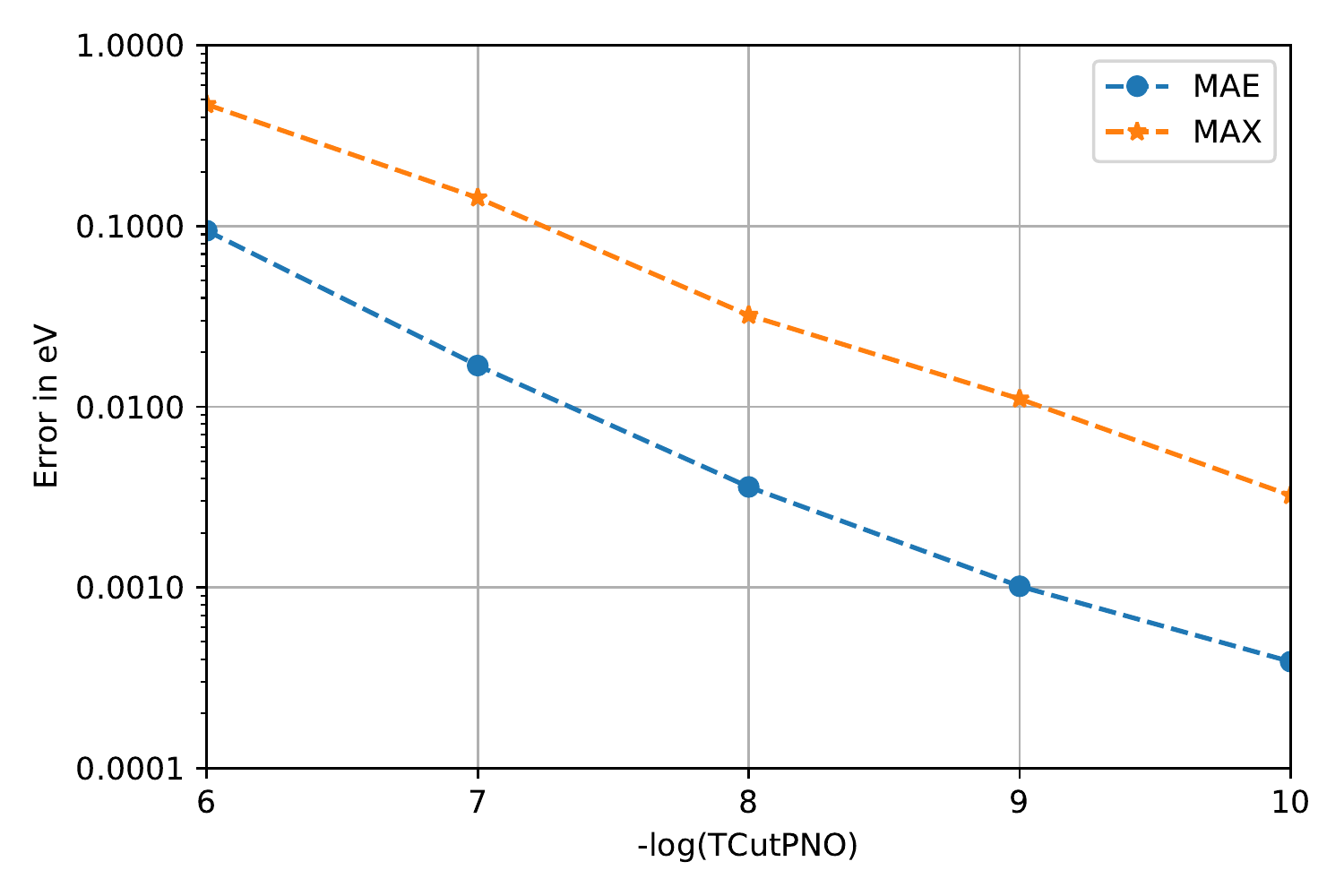}
	\caption{Mean absolute (MAE) and maximum (MAX) PNO truncation errors (in eV) of PNO-EOM-CCSD/aug-cc-pVTZ excitation energies for the 28-molecule benchmark set.}
	\label{fig:eom:benchmark_atz_ee}
\end{figure}

\begin{figure}
	\centering
	\includegraphics[width=0.7\linewidth]{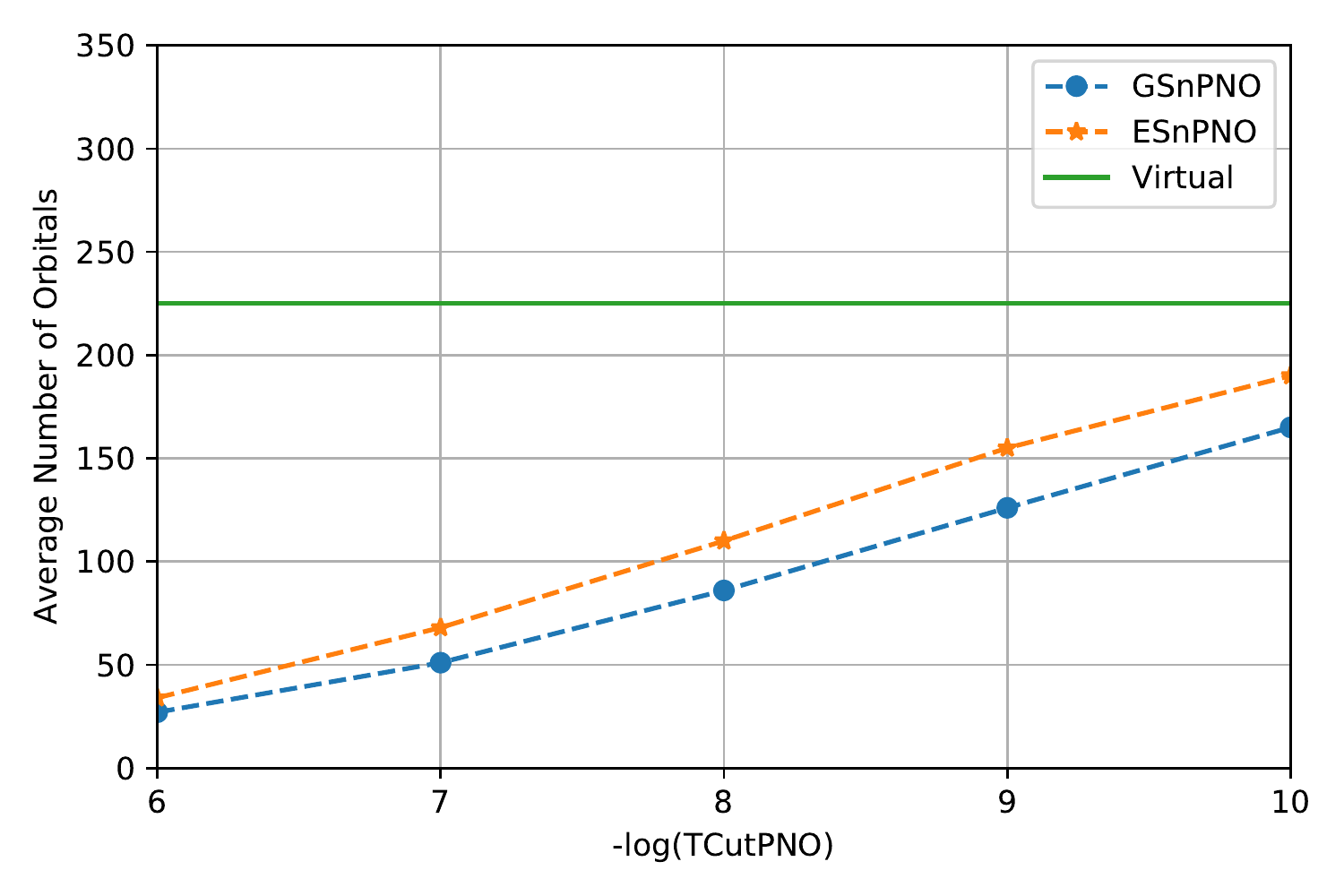}
	\caption{ Convergence of average number of PNOs per pair per molecule in ground state (GSnPNO) and excited states (ESnPNO) of PNO-EOM-CCSD/cc-pVTZ }
	\label{fig:eom:benchmark_tz_pno}
\end{figure}

\begin{figure}
	\centering
	\includegraphics[width=0.7\linewidth]{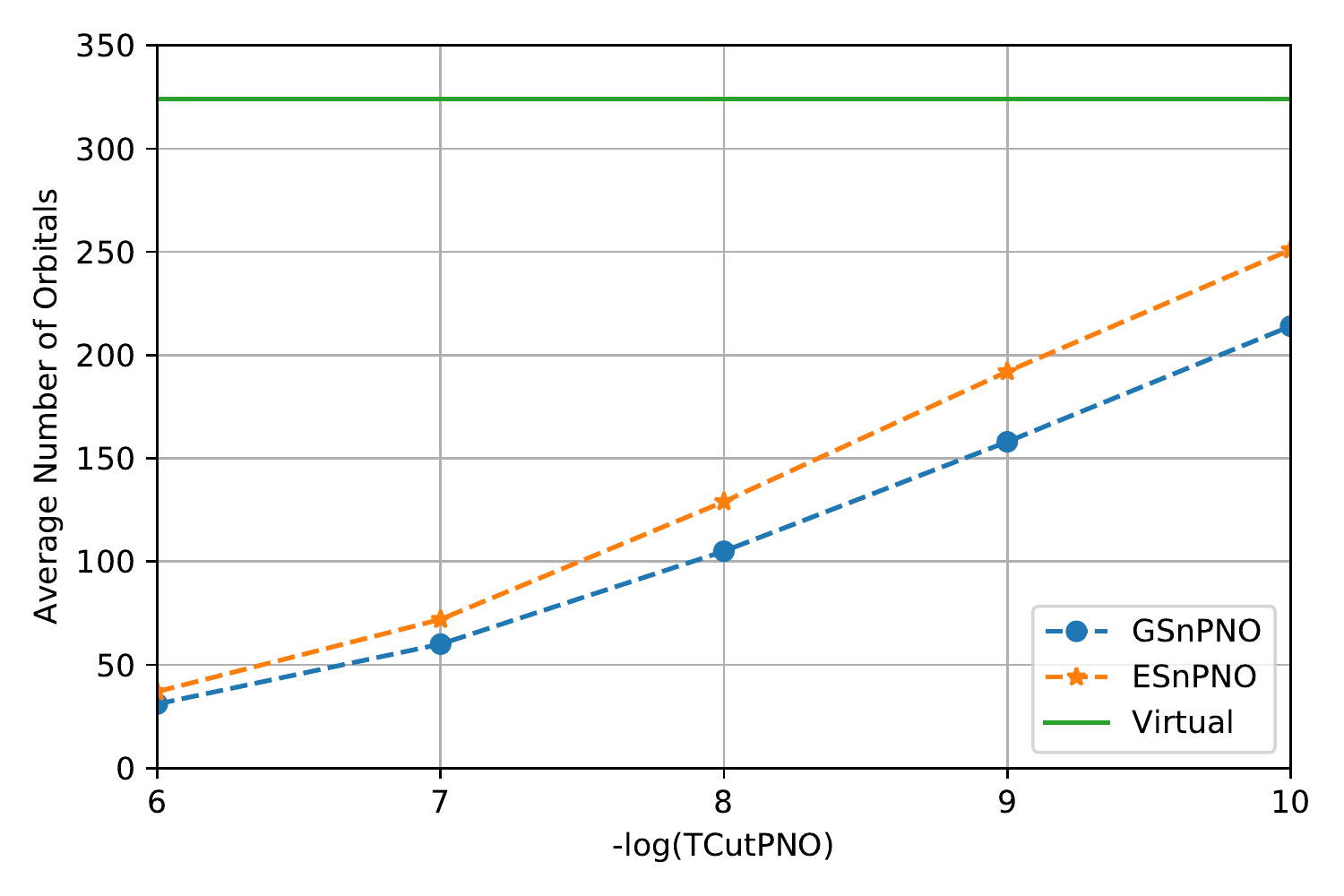}
	\caption{ Convergence of average number of PNOs per pair per molecule in ground state (GSnPNO) and excited states (ESnPNO) of PNO-EOM-CCSD/aug-cc-pVTZ }
	\label{fig:eom:benchmark_atz_pno}
\end{figure}

To further test the performance of state-averaged PNOs, we used the PNO-EOM-CCSD method to compute the excitation energies of the lowest six singlet excited states of 28 organic molecules in the benchmark dataset from Thiel \etal \cite{Schreiber2008}.
Variation of statistical measures of the errors with the $T_{\textmd{CutPNO}}$ parameter are presented in Fig. \ref{fig:eom:benchmark_tz_ee} and Fig. \ref{fig:eom:benchmark_atz_ee} for cc-pVTZ and aug-cc-pVTZ basis sets, respectively. The corresponding average numbers of ground-state and excited-state PNOs are shown in Fig. \ref{fig:eom:benchmark_tz_pno} and Fig. \ref{fig:eom:benchmark_atz_pno}, respectively.

The average excitation energy errors become smaller than 0.1 eV already with  $T_{\textmd{CutPNO}}$=$10^{-6}$, further reduce to below 0.02 eV with $T_{\textmd{CutPNO}}$=$10^{-7}$, and further decrease monotonically with $T_{\textmd{CutPNO}}$ for both basis sets. The maximum errors also decrease monotonically, but require tighter truncation, $T_{\textmd{CutPNO}}$=$\{10^{-7},10^{-8}\}$ for the \{cc-pVTZ,aug-cc-pVTZ\} basis set, to reduce below 0.1 eV.
Hence, a threshold of $10^{-7}$ is suitable for general applications while a threshold of $10^{-8}$ should be sufficient for high-accuracy applications.
With these thresholds, the average numbers of ground/excited state PNOs are $\sim$50/70 and $\sim$100/120 for cc-pVTZ and aug-cc-pVTZ basis, respectively, which is a significant decrease over the average number of total virtual orbitals. For all values of $T_{\textmd{CutPNO}}$ the number of excited state PNOs is only 20\%-30\% higher than the corresponding
number of ground state PNOs.

\subsection{Rydberg and Charge Transfer States}
To study the accuracy of state-averaged PNO-EOM-CCSD on excited states with Rydberg and charge transfer character we selected two prototypical examples: states $S_1$ and $S_4$ of acetamide (Table \ref{table:eom:acetamide}) and states $S_1$ and $S_2$ of the ethylene-tetrafluoroethylene ($\ce{C2H4}-\ce{C2F4}$) model\cite{Dreuw2003} (Table \ref{table:eom:c2h4_c2f4}). The latter model was also used to test other PNO-based excited state methods,
by H\"attig and Helmich \cite{Helmich2013} and by Dutta \etal \cite{Dutta2016}.

\begin{table}[!h]
	\caption{Truncation errors ( eV) of the aug-cc-pVTZ \textsuperscript{\emph{a}} PNO-EOM-CCSD excitation energies of the four lowest singlet states of acetamide . States $S_1$ and $S_4$ have strong Rydberg character. Excited-state PNOs were averaged over lowest 10 states.}
	\begin{center}
		\begin{tabular}{lrrrrr}
			\toprule
			$T_{\textmd{CutPNO}}$ &    $10^{-6}$&     $10^{-7}$ &     $10^{-8}$ &     $10^{-9}$ &    $10^{-10}$ \\
			ESnPNO &          32  &          69   &        125    &       191     &       245     \\
			\midrule
			$S_1$     &        0.0699 &        0.0052 &       -0.0017 &       -0.0009 &       -0.0002 \\
			$S_2$     &        0.0038 &       -0.0058 &       -0.0032 &       -0.0012 &       -0.0005 \\
			$S_3$     &        0.0185 &       -0.0001 &       -0.0028 &       -0.0011 &       -0.0003 \\
			$S_4$     &        0.0464 &        0.0012 &       -0.0023 &       -0.0004 &        0.0000 \\
			\bottomrule
		\end{tabular}
		\label{table:eom:acetamide}
	\end{center}    
	\begin{tablenotes}
		\item[1] \textsuperscript{\emph{a}} Total number of unoccupied orbitals = 283.
	\end{tablenotes}
	
\end{table}

The truncation errors for the two Rydberg states ($S_1$ and $S_4$) were found to be somewhat larger than the errors for the valence states ($S_2$ and $S_3$) with $T_{\textmd{CutPNO}}$=$10^{-6}$ but they are comparable with $T_{\textmd{CutPNO}}$=$10^{-7}$ or tighter.
Overall, no significant differences in the performance of excited state PNOs is observed for Rydberg and non-Rydberg states.
\begin{table}[!h]
	\caption{Truncation errors ( eV) of the aug-cc-pVDZ \textsuperscript{\emph{a}} PNO-EOM-CCSD excitation energies of the four lowest singlet states of the \ce{C2H4}-\ce{C2F4} dimer separater by 10 a.u. (Ref. \citenum{Dreuw2003}). States $S_1$ and $S_2$ have charge-transfer character. Excited-state PNOs were averaged over lowest 10 states.}
	\begin{center}
		\begin{tabular}{lrrrrr}
			\toprule
			$T_{\textmd{CutPNO}}$ &    $10^{-6}$&     $10^{-7}$ &     $10^{-8}$ &     $10^{-9}$ &    $10^{-10}$ \\
			ESnPNO &          8   &         19    &        37     &       63      &       93      \\    
			\midrule
			$S_1$     &        0.1330 &        0.0321 &        0.0087 &        0.0022 &        0.0006 \\
			$S_2$     &        0.2059 &        0.0513 &        0.0100 &        0.0014 &        0.0001 \\
			$S_3$     &        0.0153 &        0.0013 &        0.0004 &        0.0003 &        0.0001 \\
			$S_4$     &        0.0164 &        0.0005 &       -0.0001 &       -0.0000 &        0.0000 \\
			\bottomrule
		\end{tabular}
		\label{table:eom:c2h4_c2f4}
	\end{center}

	\begin{tablenotes}
	\item[1] \textsuperscript{\emph{a}} Total number of unoccupied orbitals = 188.
	\end{tablenotes}
	
\end{table}

The truncation errors for the two charge transfer states were found to be substantially larger than the valence states at all truncation thresholds.
$T_{\textmd{CutPNO}}$=$10^{-7}$ was required to reduce the errors to below 0.1 eV, and $T_{\textmd{CutPNO}}$=$10^{-8}$ is sufficient to reduce the errors to
0.01 eV for charge transfer excitations.
This is in agreement with the findings of H\"attig and Helmich \cite{Helmich2011}, who pointed out that it requires more PNOs to get the same accuracy for charge transfer excitations. 
They attributed this to the use of the semicanonical CIS(D) amplitudes in constructing the excited-state PNOs (Eq. \eqref{eq:cis_d}); with localized occupied orbitals the off-diagonal matrix elements of the Fock operator are substantial and cannot be neglected. Nevertheless, the performance of semicanonical PNOs is still acceptable.

\section{Conclusions}
We proposed a state-averaged PNO ansatz for efficient and simple treatment of manifolds of excited states in the context of reduced-scaling
excited-state many-body methods. We evaluated the performance of the state-averaged PNO ansatz
in the context of PNO-EOM-CCSD method for prediction of excitation energies. The PNO-EOM-CCSD implementation
is based on a new massively parallel canonical implementation of EOM-CCSD in the MPQC program.
The state-averaged PNO-EOM-CCSD approach has been tested on the first six excited states of 28 organic molecules, yielding an average truncation error below 0.020 eV at $T_{\textmd{CutPNO}}$=$10^{-7}$ for both the cc-pVTZ and aug-cc-pVTZ basis sets.
With this truncation threshold, the number of state-averaged PNOs is reduced by more than 70\% for cc-pVTZ and 80\% for aug-cc-pVTZ.
Overall, the state-averaged PNOs provide excellent accuracy for low lying valence and Rydberg states, but more PNOs are required to achieve the same accuracy for charge transfer states.
These results are sufficiently encouraging to warrant the development of a production-level PNO-EOM-CCSD code based on the state-averaged PNO definitions introduced here.
	
\bibliography{pno_eom_ccsd}
	
\end{document}